\documentclass[a4paper,fleqn,usenatbib]{mnras}

\usepackage{newtxtext,newtxmath}

\usepackage[T1]{fontenc}
\usepackage{ae,aecompl}

\usepackage{graphicx}	
\usepackage{amsmath}	
\usepackage{amssymb}	

\title[The X-ray faint galaxy group MRC\,0116+111]{Magnetic fields and extraordinarily bright radio emission in the X-ray faint galaxy group MRC\,0116+111}

\author[F. Mernier et al.]{
F. Mernier,$^{1,2,3}$\thanks{E-mail: mernier@caesar.elte.hu}
N. Werner,$^{1,4,5}$
J. Bagchi,$^{6}$
A. Simionescu,$^{3,7,8}$
H. B{\"o}hringer,$^{9}$
\newauthor
S. W. Allen$^{10,11,12}$
and J. Jacob$^{13}$
\\
$^{1}$MTA-E\"otv\"os University Lend\"ulet Hot Universe Research Group, P\'azm\'any P\'eter s\'et\'any 1/A, Budapest, 1117, Hungary\\
$^{2}$Institute of Physics, E\"otv\"os University, P\'azm\'any P\'eter s\'et\'any 1/A, Budapest, 1117, Hungary\\
$^{3}$SRON Netherlands Institute for Space Research, Sorbonnelaan 2, NL-3584 CA Utrecht, The Netherlands\\
$^{4}$Department of Theoretical Physics and Astrophysics, Faculty of Science, Masaryk University, Kotl\'a\v{r}sk\'a 2, Brno, CZ-611 37, Czech Republic \\
$^{5}$School of Science, Hiroshima University, 1-3-1 Kagamiyama, Higashi-Hiroshima 739-8526, Japan \\
$^{6}$Inter-University Centre for Astronomy and Astrophysics, Post Bag 4, Ganeshkhind, Pune 411007, India\\
$^{7}$Leiden Observatory, Leiden University, PO Box 9513, NL-2300 RA Leiden, The Netherlands\\
$^{8}$Kavli Institute for the Physics and Mathematics of the Universe (WPI), University of Tokyo, Kashiwa 277-8583, Japan\\
$^{9}$Max-Planck-Institut f\"ur extraterrestrische Physik, Giessenbachstrasse, Garching D-85748, Germany\\
$^{10}$Department of Physics, Stanford University, 382 Via Pueblo Mall, Stanford, CA 94305, USA\\
$^{11}$SLAC National Accelerator Laboratory, 2575 Sand Hill Road, Menlo Park, CA 94025, USA\\
$^{12}$Kavli Institute for Particle Astrophysics and Cosmology, Stanford University, 452 Lomita Mall, Stanford, CA 94305, USA\\
$^{13}$Newman College, Thodupuzha, Kerala, 685584, India
}

\date{Accepted 2019 April 30. Received 2019 April 5; in original form 2019 February 25}

\pubyear{2018}

\begin{document}
\label{firstpage}
\pagerange{\pageref{firstpage}--\pageref{lastpage}}
\maketitle

\begin{abstract}
MRC\,0116+111 is a nearby ($z=0.132$) poor galaxy group, which was previously known for exhibiting a bright diffuse radio emission with no central point-like source, presumably related to a past activity of the active galactic nucleus (AGN) in its central cD galaxy. Here, we present an X-ray observation ($\sim$30 ks of cleaned \textit{XMM-Newton}/EPIC exposure) of this system, allowing us for the first time a detailed comparison between the thermal and non-thermal components of its intragroup medium (IGrM). Remarkably, we find that the radio-to-X-ray luminosity ratio is among the highest ever observed for a diffuse extragalactic source so far, while the extent of the observed radio emission is about three times larger than its observed soft X-ray emission. Although powerful AGN activity may have disturbed the dynamics of the thermal IGrM in the form of turbulence, possibly re-energizing part of the relativistic electron population, the gas properties lie within the $L_X$--$T$ scaling relation established previously for other groups. The upper limit we find for the non-thermal inverse-Compton X-ray emission translates into a surprisingly high lower limit for the volume-averaged magnetic field of the group ($\ge$4.3 $\mu$G). Finally, we discuss some interesting properties of a distant ($z \simeq 0.525$) galaxy cluster serendipitously discovered in our EPIC field of view.
\end{abstract}





\section{Introduction}\label{sect:intro}


Well beyond their radii of gravitational influence, supermassive black holes (SMBH) are expected to play a fundamental role in the formation and evolution of the largest structures of the Universe. While the tight correlation between black hole mass and stellar velocity dispersion of the bulge of their galaxy hosts strongly suggests that the formation and growth of these two components happened together \citep{ferrarese2000,gebhardt2000,kormendy2013}, active galactic nuclei (AGN) in the centre of massive elliptical galaxies often produce powerful jets -- or lobes -- that are visible at radio wavelengths \citep[e.g.][]{kataoka2005}. This form of AGN feedback (often named as 'kinetic" or "radio" mode, by contrast to the "radiative" or "quasar" mode that operates when the SMBH is close to the Eddington limit) is thought to play a crucial role in regulating the thermal balance of the hot, X-ray emitting atmospheres of massive ellipticals, groups and clusters of galaxies, as these radio jets provide enough mechanical energy to offset the rapid cooling of the gas and to prevent the formation of new stars, thus keeping giant ellipticals "red and dead" \citep[for recent reviews, see][]{peterson2006,fabian2012,mcnamara2012,werner2019}. In cool-core systems\footnote{Contrary to non-cool-core systems, the cooling time of the hot atmosphere at the centre of cool-core systems is typically less than the Hubble time. Consequently, these (mostly relaxed) systems have a centrally peaked surface brightness coupled with positive temperature and entropy gradients \citep[see e.g.][]{molendi2001,hudson2010}.}, specifically, AGN are thought to inject lobes of relativistic plasma and magnetic fields in the intracluster medium (ICM), in the intragroup medium (IGrM), and in hot atmospheres of elliptical galaxies, thereby creating cavities that are easily observed with the current generation of X-ray observatories \citep[e.g.][]{fabian2000,mcnamara2000,wise2007,randall2011,hlavacek2012}. 

How exactly these bubbles -- filled by relativistic, non-thermal particles and magnetic fields -- detach, rise buoyantly and re-heat isotropically the surrounding ICM (IGrM) to prevent massive "cooling flows" is still unclear. Among the difficulties encountered in both observations and simulations, our knowledge of magnetic fields in clusters and groups -- in particular their origin(s), topologies, coupling with AGN feedback, and even their average intensities -- is rather limited. In fact, the existence of magnetic fields in the ICM (IGrM) is revealed because of the synchrotron radio emission radiated by accelerating non-thermal electrons along magnetic field lines \citep[for a review, see e.g.][]{bruggen2012}. In addition to the jets/lobes tracing the most recent AGN activity, synchrotron emission is observed under essentially three distinct radio morphologies \citep{ferrari2008,feretti2012,vanweeren2019}:
\begin{enumerate}
\item Giant radio haloes in merging clusters ($\gtrsim$1 Mpc), presumably tracing an old population of non-thermal electrons that has been recently re-accelerated during the merger;
\item Radio mini-haloes, found only in the inner $\lesssim$500 kpc of cool-core clusters, where an old population of non-thermal electrons has been possibly re-accelerated by AGN feedback;
\item Radio relics, which exhibit an elongated morphology sometimes extending to Mpc scale, presumably tracing shocks in the outskirts of merging clusters \citep[e.g.][]{bagchi2006}.
\end{enumerate}
However, radio emission alone cannot constrain directly magnetic field intensities because the synchrotron emission also depends on the (\textit{a priori} unknown) relativistic electron density. One approximation commonly found in the literature to break this degeneracy is to assume that the total (i.e. magnetic and particle) energy density in the relativistic plasma is minimal, which implies that the contributions from magnetic and particle energy densities are roughly equal. This so-called "equipartition" assumption provides magnetic field estimates of the order of 0.1--1 $\mu$G in radio haloes \citep{petrosian2008,feretti2012} and a few tens of $\mu$G in central radio lobes/mini-haloes \citep[e.g.][]{birzan2008}. Because particles and magnetic fields do not necessarily have the same origin and history, however, the validity of the equipartition approximation is not obvious. Magnetic field intensities can also be estimated via Faraday rotation measurements \citep[e.g.][]{carilli2002,govoni2004,bohringer2016}. Although the derived magnetic fields are globally of the same order of magnitude as when using the equipartition assumption \citep[e.g.][and references therein]{govoni2010}, there may be appreciable differences due to the fact that Faraday rotation measurements are integrated along the line of sight, and depend thus on the local magnetic field topology (and the possible foreground contamination).

In principle, non-thermal electrons responsible for the radio emission should also boost the energy of the cosmic microwave background (CMB) photons up to keV units via inverse-Compton (IC) scattering, hence producing an additional non-thermal emission in the X-ray band \citep{feenberg1948,felten1966,harris1979,rephaeli1979,schlickeiser1979,rephaeli1988,bagchi1998}. Interestingly, since the IC emissivity depends on the relativistic electron population but not on the magnetic field, simultaneous radio synchrotron and IC measurements allow to determine the volume-averaged magnetic field in clusters and groups without the need of the equipartition assumption \citep{bagchi1998}. So far, this method has provided essentially lower limits on magnetic field intensities of clusters hosting giant haloes or relics \citep[][and references therein]{bartels2015} simply because only upper limits on the IC emission could be constrained. The situation is more problematic for mini-haloes and central radio lobes in cool-core systems, since the emission of the 0.3--10 keV band is almost entirely dominated by the (centrally peaked) thermal X-ray photons, making the non-thermal IC emission virtually impossible to detect. With the current X-ray telescopes, robust constraints on the volume-averaged ICM/IGrM magnetic field within the zone of influence of the central AGN feedback could be inferred only for systems that would be exceptionally bright in radio and with very limited soft (i.e. thermal) X-ray luminosities.

In this paper, we focus on the X-ray emission of the galaxy group MRC\,0116+111 (hereafter, MRC\,0116). Initially discovered by the Ooty Lunar Occultation Survey at 327 MHz \citep{joshi1980}, this radio source -- also member of the Molonglo Reference Catalogue \citep{large1981} -- had been reported a first time using the \textit{Giant Metrewave Radio Telescope} (\textit{GMRT}) and the \textit{Very Large Array} (\textit{VLA}) data \citep{gopal2002}, then discussed in detail with more recent \textit{GMRT} and optical observations \citep[][where the authors find the dominant galaxy to be at $z=0.1316$]{bagchi2009}. Although, based on optical observations, the source had also been firmly classified as a galaxy cluster \citep{lopes2004}, the limited number of galaxies seen in the optical band \citep{bagchi2009} suggests instead a rather poor galaxy group. Although, interestingly, the radio source was reported with a surprisingly high luminosity ($L_\text{621 MHz} \sim 1.21 \times 10^{25}$ W\,Hz$^{-1}$ and $L_\text{1.4 GHz} \sim 4.57 \times 10^{24}$ W\,Hz$^{-1}$) and with a mini-halo-like morphology (without emission from the central SMBH), no X-ray observation of MRC\,0116 has been reported so far in the literature. In fact, results from the \textit{ROSAT} All Sky Survey are consistent with no detection at the position of the source \citep{boller2016}, which strongly suggests an anomalously low X-ray-over-radio luminosity ratio. In addition to be a target of potentially high interest to constrain X-ray non-thermal IC emission, this system might have witnessed an extremely powerful (past) AGN activity \citep{bagchi2009}, whose unique impact on the surrounding IGrM is worth studying. 

Here, we present for the first time an \textit{XMM-Newton}/EPIC observation of MRC\,0116. The data reduction and analysis are described in Sect.~\ref{sect:data_reduction}. Subsequent results and their interpretation are detailed and discussed in Sect.~\ref{sect:results} and Sect.~\ref{sect:discussion}, respectively. In these two sections, we also report and discuss the properties of a distant galaxy cluster serendipitously discovered in the same pointing. Finally, Sect.~\ref{sect:conclusion} summarizes our findings. Throughout this paper, we assume a $\Lambda$CDM cosmology with $\Omega_m = 0.3$, $\Omega_\Lambda = 0.7$, and $H_0 = 70$ km\,s\,Mpc$^{-1}$. At $z=0.132$, 1 arcmin corresponds to $\sim$180 kpc. All the errors given in the following correspond to their 68\% confidence level, unless stated otherwise.


\section{Data preparation}\label{sect:data_reduction}


\subsection{\textit{XMM-Newton} data}\label{sect:data_xmm}

MRC\,0116 was observed by \textit{XMM-Newton} on 2010 January 10--11 (ObsID:0722900101), with $\sim$56 ks of raw exposure. We use the latest version (v17.0.0) of the \textsc{XMM SAS} software, as well as the associated calibration files of 2018 February, to reduce the data. We first process the EPIC data using the pipeline commands \texttt{empproc} and \texttt{epproc} for the MOS (i.e. MOS\,1 and MOS\,2) and pn instruments, respectively. We then filter the data from significant soft-protons flares. We do so using the \texttt{espfilt} command on the MOS and pn light curves in the 10--12 keV band. This command builds count histograms within 100 s time bins and selects the count threshold above which the histograms deviate from a Poisson distribution by >2$\sigma$. We then apply the appropriate good time intervals on the raw data to obtain soft proton cleaned event lists. We repeat the same procedure in the 0.3--2 keV band, as some independent flares might as well be detected at softer energies \citep{lumb2002}. Unfortunately, about half of the observation was significantly affected by flaring events. The total net exposure of the MOS\,1, MOS\,2, and pn data is 28, 37, and 28 ks, respectively. 

\begin{figure}
                \includegraphics[width=0.5\textwidth,trim={0 45 0 0},clip]{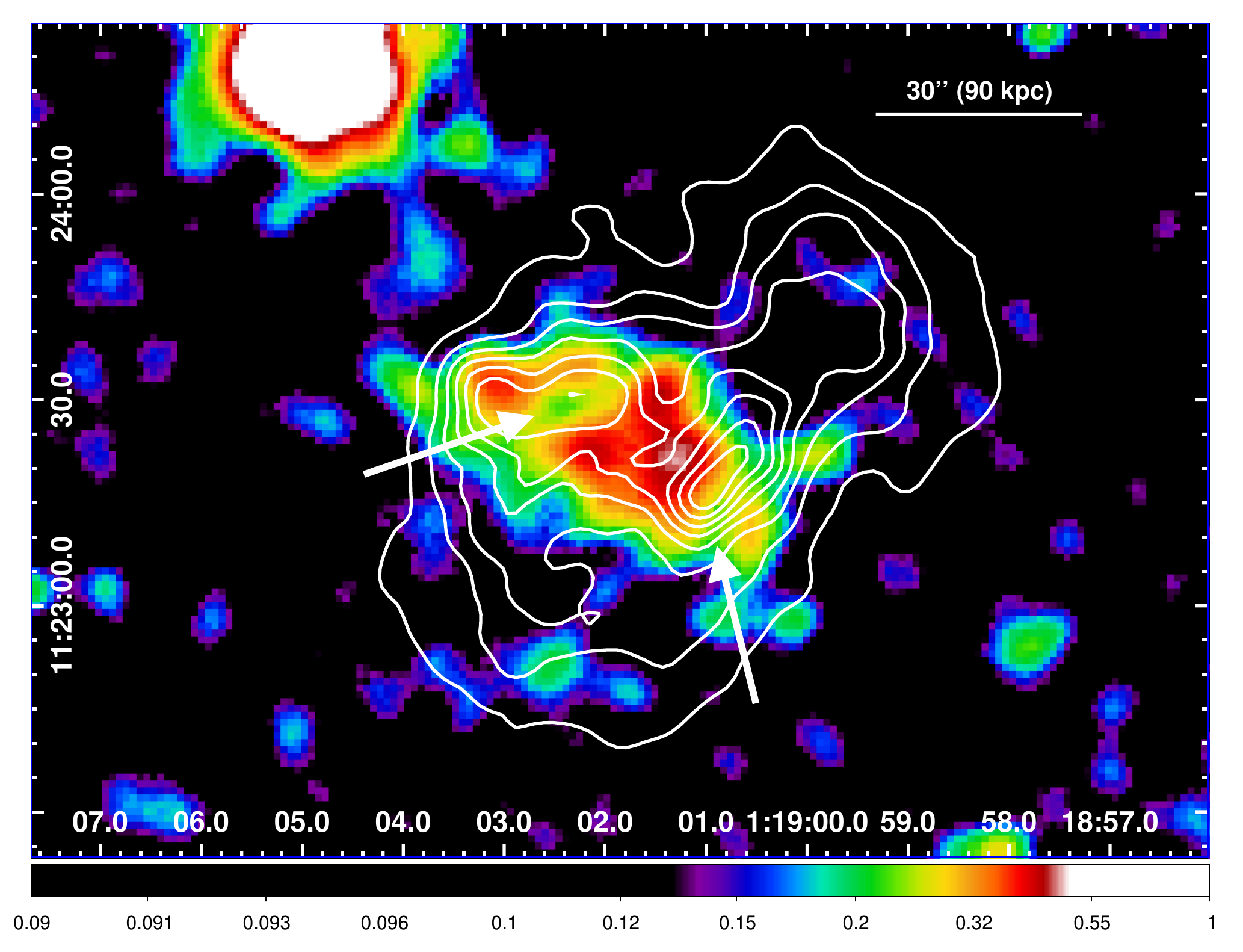}

        \caption{Background- and vignetting-corrected \textit{XMM-Newton} EPIC image of MRC\,0116+111 in the 0.3--2 keV band. The white arrows indicate the two apparent surface brightness drops, possibly associated with cavities (see text). The radio contours (\textit{GMRT}, 610 MHz) from \citet{bagchi2009} are overplotted in white. Both the X-ray and radio images have a spatial resolution of $\sim$6 arcsec FWHM.}
\label{fig:xray_radio_image}
\end{figure}

\begin{figure*}
        \centering
                \includegraphics[width=0.49\textwidth,trim={14 42 14 0},clip]{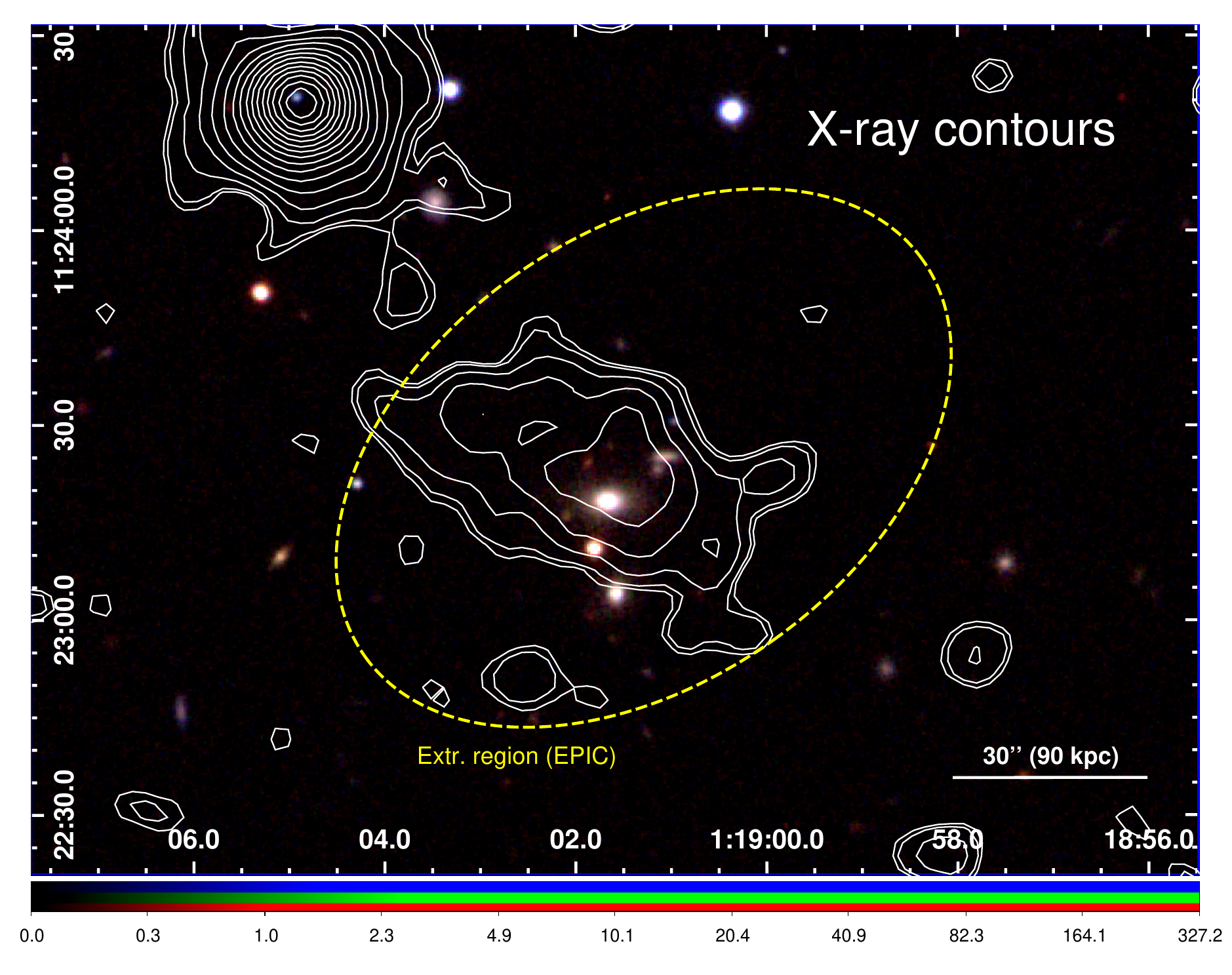}
                \includegraphics[width=0.49\textwidth,trim={14 42 14 0},clip]{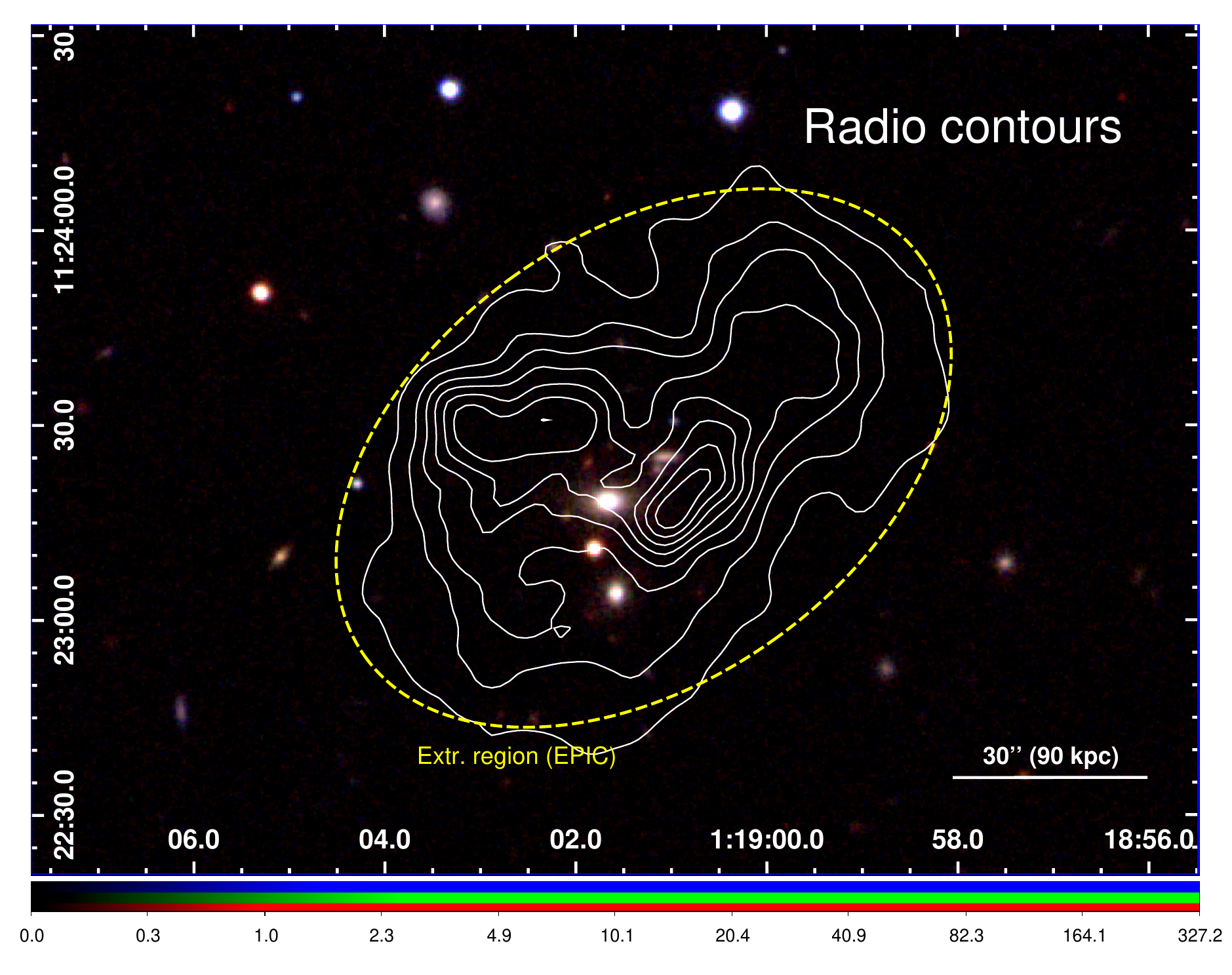} \\

        \caption{Composite optical image of MRC\,0116+111 from the SDSS-III data ($g$, $r$, and $i$ bands). The source region extracted for the spectral analysis is delimited by the yellow dashed ellipse. The X-ray contours (EPIC, 0.3--2 keV) and the radio contours (\textit{GMRT}, 610 MHz) are overplotted in the left and right panels, respectively. Both the X-ray and radio images have a spatial resolution of $\sim$6 arcsec FWHM.}
\label{fig:xray_radio_contours}
\end{figure*}

For each instrument, we generate a raw image, a background image, and an exposure map within the 0.3--2 keV band using the \textsc{SAS} task \texttt{evselect}. The background images are obtained from filter wheel closed data ($\sim$50 ks for each instrument). Before subtracting it from the raw image, we re-scale the background image accordingly to the 10--12 keV energy band, where no source emission is expected. The exposure maps are generated for each of the three instruments using the \textsc{SAS} task \texttt{eexpmap}. The three background-subtracted images and the three exposure maps are first merged separately before we divide the total EPIC image by the total EPIC exposure map with appropriate weights. The final background- and vignetting-corrected EPIC image of MRC\,0116 (in the 0.2--3 keV band) is shown in Fig.~\ref{fig:xray_radio_image}. For comparison, we overplot the radio contours (\textit{GMRT}, 610 MHz) presented in \citet{bagchi2009}. In Fig.~\ref{fig:xray_radio_contours} (left-hand and right-hand panels, respectively), we also show the X-ray and radio contours separately, overplotted on an optical/infrared mosaic using the $g$, $r$, and $i$ bands from the SDSS-III data \citep{eisenstein2011}.

Regrettably, the limited total number of net counts ($\sim$415) detected from MRC\,0116 and the moderate spatial resolution of \textit{XMM-Newton} do not allow us to investigate its spatial structure in detail via spectroscopy. However, we can reasonably perform a spectral analysis within the full extent of the system. We extract the MOS\,1, MOS\,2, and pn spectra within an ellipse region encompassing the bulk of the radio emission (Fig.~\ref{fig:xray_radio_contours}, the yellow dashed contour). After visual inspection and using the \textsc{SAS} task \texttt{edetect\_chain}, no point-like sources were detected in this region (see also Sect.~\ref{sect:pointsources}). The redistribution matrix file (RMF) and the ancillary response file (ARF) are computed by using the commands \texttt{rmfgen} and \texttt{arfgen}, respectively. Throughout this paper, the spectral analysis is done using the \textsc{SPEX} (v3.04) fitting package \citep{kaastra1996,kaastra2017b}. We consequently convert the raw spectra, the RMFs, and the ARFs to \textsc{SPEX} readable files using the appropriate subroutine \texttt{trafo}. 

The rather low source-to-background count rate ratios, found to be $\sim$1.1 and $\sim$0.34, respectively, in the 0.3--2 keV band (imaging analysis, Sect.~\ref{sect:image}) and the 0.6--5 keV band (spectral analysis, Sect.~\ref{sect:spectral}), stress the importance of a careful estimate of the background when analysing X-ray spectra of faint, diffuse objects. In the case of nearby, extended clusters, the entire field of view is often contaminated by the emission of the source, and the safest approach requires to properly model all the background components \citep[see e.g.][and references therein]{mernier2015}. Here, however, the very limited angular extent of MRC\,0116 (i.e. less than 1 arcmin) allows us to estimate the background directly from a nearby, individual region in the same field of view. All the spectral components from this local background region are expected to behave in a very similar way as along the line of sight of the source, and its spectra can be thus directly subtracted from the raw spectra of the corresponding source region. We therefore select as local background a box region located on the same MOS\,1, MOS\,2, and pn chip as the source (within 2.8 arcmin or less from it), about three times larger than the source extraction region, and in which point-like sources have been removed. The subtracted spectra, showing net counts only, can be then fitted without background modelling. As demonstrated in Appendix~\ref{sect:bg_uncertainties}, modelling the background (instead of subtracting it) does not affect significantly the results presented in the following sections, hence the present method can be considered as robust.

\subsection{\textit{Chandra} data}\label{sect:data_chandra}

In addition to \textit{XMM-Newton}, MRC0116 was observed with \textit{Chandra}/ACIS on 2010 June 24 (ObsID:11865). These data are reduced using the dedicated software \textsc{CIAO} (v4.9). After reprocessing the data following the recommended procedure (i.e. via the \texttt{chandra\_repro} command), we create an exposure-corrected ACIS image in the full band of the instrument (0.5--7 keV) using the task \texttt{fluximage}. The very limited exposure of this pointing (18 ks, resulting in only 73 net counts for a signal-to-background ratio of $\sim$0.64 in the 0.3--2 keV band) prevents us from deriving any spatial (and thus spectral) information of the extended emission; therefore the rest of the paper will be essentially devoted to the analysis of the \textit{XMM-Newton} observation. Nevertheless, the ACIS data can be used to put useful constraints on the emission from individual point-like sources within this group (Sect.~\ref{sect:pointsources}).


\section{Results}\label{sect:results}


\subsection{Imaging analysis}\label{sect:image}

The optical image of MRC\,0116 suggests that this group does not contain more than a few galaxy members (see also Sect.~\ref{sect:intro}). As shown in Fig.~\ref{fig:xray_radio_contours} (left), the location of the brightest group galaxy, USNO-A2.0\,0975-00295178 \citep{monet1998}, is formally consistent with the location of the X-ray peak (less than $\sim$6 arcsec, which is also the point spread function of the EPIC instruments). However, Figs. \ref{fig:xray_radio_image} and \ref{fig:xray_radio_contours} (left) suggest that the IGrM is disturbed. This can be further quantified by computing the centroid shift parameter $w$ \citep[as defined in][]{mohr1995,rasia2013}. Using 10 circular regions centred of the X-ray surface brightness peak with their aperture radii spanning from 6 arcsec to 1 arcmin, we find $w \simeq 0.035$, which is typical for a disturbed system \citep{cassano2010}. 

Another striking feature seen from Fig.~\ref{fig:xray_radio_image} is the apparent surface brightness drop seen at $\sim$40--50 kpc north-east from the X-ray peak, beyond which an elongated re-enhancement seems to follow the radio contours of the eastern lobe. Although the small number of counts prevents us from confirming this feature with sufficient significance, this might suggest the existence of an X-ray cavity at this location. This possibility will be further discussed in Sect.~\ref{sect:AGN_activity}. Though even less significant, a similar X-ray decrement, possibly corresponding to another cavity, may be seen $\sim$30--40 kpc south-west from the X-ray peak (roughly surrounding the western radio lobe). These two apparent surface brightness drops are indicated by the white arrows on Fig.~\ref{fig:xray_radio_image}.

The net number of counts obtained for the source in the 0.3--2 keV band of the MOS\,1, MOS\,2, and pn instruments is 88, 126, and 201, respectively. At first approximation, X-ray fluxes (and luminosities) can be calculated directly from the count rate inferred from our EPIC images. Using the web tool WebPIMMS\footnote{https://heasarc.gsfc.nasa.gov/cgi-bin/Tools/w3pimms/w3pimms.pl} and assuming a gas temperature of $kT = 0.76$ keV (Sect.~\ref{sect:spectral}), we find that the fluxes and luminosities estimated within the 0.5--7 keV band from our MOS\,1, MOS\,2, and pn individual images are all $<$2$\sigma$ consistent with the flux and luminosity $f_\text{X,0.5--7}$ and $L_\text{X,0.5--7}$, estimated from the spectral analysis and reported in Table~\ref{table:fitting_results}.

\subsection{Spectral analysis}\label{sect:spectral}

\begin{figure}
        \centering
                \includegraphics[width=0.49\textwidth]{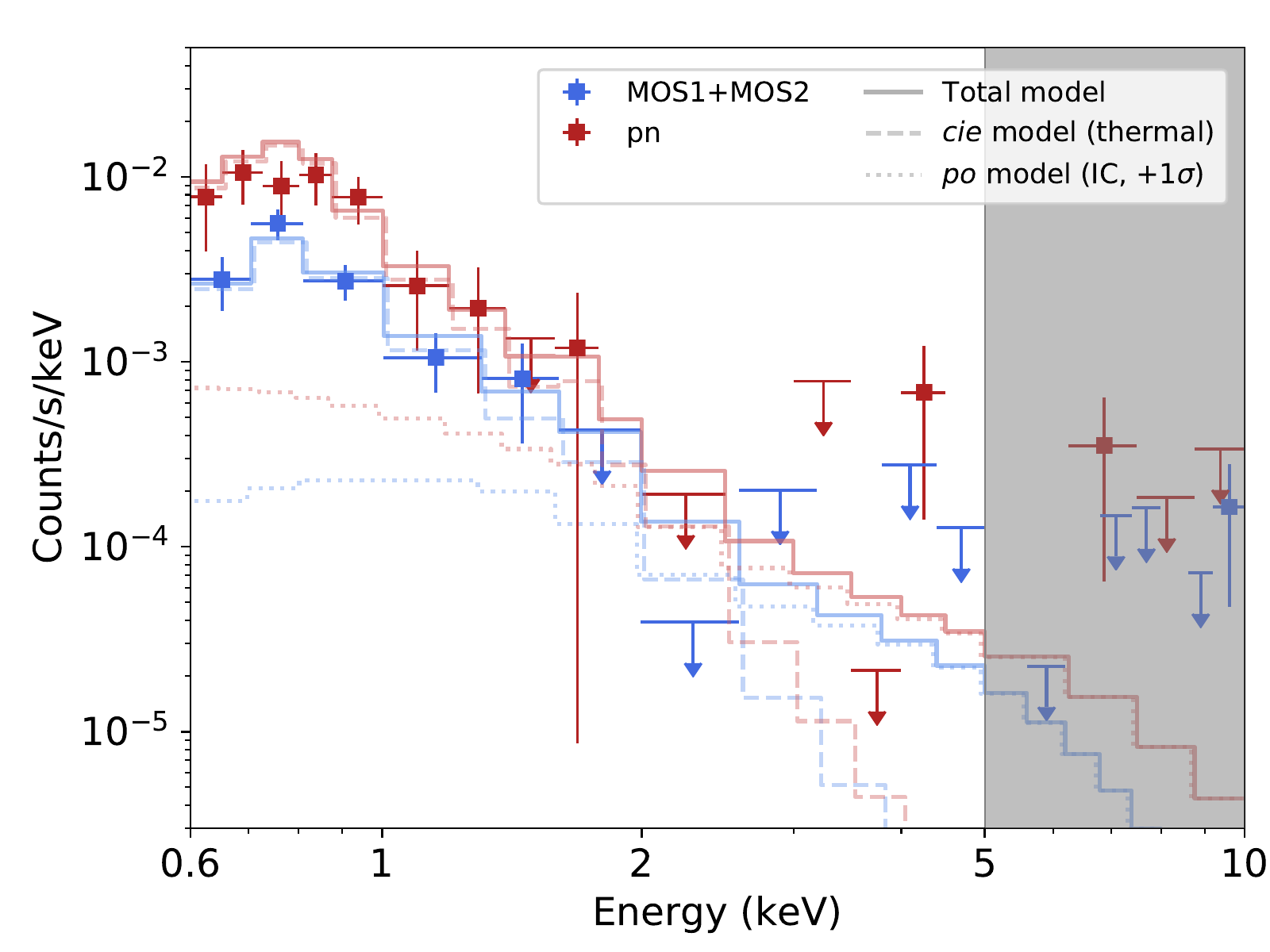}

        \caption{Background subtracted \textit{XMM-Newton} EPIC spectra of MRC\,0116+111. The MOS (MOS\,1+MOS\,2) and pn spectra are fitted simultaneously with two components: (i) a \texttt{cie} (thermal) model and (ii) a (non-thermal) power law of fixed photon index $\Gamma = 1.55$, shown here on its 1$\sigma$ upper-limit. Although for display we show the observed spectra out to 10 keV, we choose to fit it only within 0.6--5 keV (see text).}
\label{fig:spectra}
\end{figure}

The background-subtracted EPIC spectra are shown in Fig.~\ref{fig:spectra}. The MOS\,1 and MOS\,2 spectra are stacked for display purpose, however they are fitted simultaneously (with tied parameters) in our analysis. In order to avoid possible biases due to gain calibration of the oxygen edge, the soft energies $E<0.6$ keV are discarded from the rest of the analysis. After grouping the channels in wider energy bins, nearly all the net counts detected by MOS and pn beyond $\sim$2 keV are consistent with zero. A few specific bins at high energies (particularly at $E>5$ keV), however, are significantly brighter (with no overlap between MOS and pn) and their reliability may be questioned as instrumental effects cannot be excluded. Therefore, and to be as conservative as possible, we choose to fit our spectra within the 0.6--5 keV band. We also verify that extending our fitting range to 10 keV does not affect the results presented throughout this paper.

In Fig.~\ref{fig:spectra}, we also note the presence of a peak of emission around $\sim$0.6--0.9 keV ($\sim$0.7--1 keV rest frame), which is characteristic of the (unresolved) Fe-L complex emitted by a low-temperature plasma (typically $kT \lesssim 2$ keV). We start by fitting simultaneously the MOS\,1, MOS\,2 and pn spectra with a redshifted ($z=0.132$; Sect.~\ref{sect:intro}) and absorbed \citep[$n_\text{H} = 3.81 \times 10^{20}$ cm$^{-2}$;][]{kalberla2005} \texttt{cie} model. This model assumes a single-temperature thermal plasma (see the \textsc{SPEX} manual for more details)\footnote{Admittedly, the ICM/IGrM usually hosts a complex temperature structure \citep[e.g.][]{frank2013,hitomi2018} and may be affected by projection effects; meaning that their emitting plasma is unlikely to be in a pure single-temperature phase. Although \textsc{SPEX} offers several multitemperature plasma models, the quality of our spectra is not sufficient to obtain further accurate information on the temperature structure of MRC\,0116.}. The free parameters are the emission measure ($Y_\texttt{cie} = \int n_e n_\text{H} dV$) and the temperature ($kT$) of the plasma. The chemical abundances, given with respect to the proto-solar units of \citet{lodders2009}, are assumed to be 0.3 \citep[e.g.][]{urban2017}. Other choices of the abundances do not affect the results of this paper. The fitting results, as well as the estimated flux and luminosity in the 0.5--7 keV band ($f_\text{X,0.5--7}$ and $L_\text{X,0.5--7}$, respectively) and the estimated X-ray bolometric luminosity ($L_\text{X,bol}$), are shown in Table~\ref{table:fitting_results} (first row). When comparing the observed C-stat (47.3) with the expected C-stat and its variance ($38 \pm 9$) obtained following the method of \citet{kaastra2017a}, we find that the model describes the data very well. The best-fitting temperature is found for $kT = (0.76 \pm 0.07)$ keV, which is typical for a poor galaxy group. Restricting the fitting range to $\sim$0.6--2 keV for safety check, we verify that this best-fitting temperature is not affected by counts at higher energies.

Since this source exhibits a high radio luminosity compared to its X-ray luminosity (see also Sect.~\ref{sect:AGN_activity}), it is an ideal target to search for IC emission in its X-ray spectrum. This is particularly relevant beyond $\sim$2 keV, where negligible thermal emission is expected from this low-temperature system. We first test the unlikely possibility that the X-ray emission would be predominantly produced by IC scattering via the relativistic electron population. This is done by modelling a (redshifted and absorbed) power-law spectrum (hereafter the \texttt{po} model) instead of a \texttt{cie} model. The parameters of interest of the \texttt{po} model are the normalisation ($Y_\texttt{po}$) and the photon index ($\Gamma$). The spectral index $\alpha_\text{X}$ of the (X-ray) IC emission is expected be the same as the spectral index $\alpha_\text{syn}$ of the (radio) synchrotron emission. \citet{bagchi2009} measured $\alpha_\text{syn} = 0.55$ for $\nu \lesssim 400$ MHz and $\alpha_\text{syn} = 1.35$ for $\nu \gtrsim 400$ MHz. Since the Lorentz factor of an electron should be on average $\gamma \gtrsim 1500$ to scatter a CMB photon up to energies of $\gtrsim 2$ keV, the radio synchrotron frequencies to be associated with X-ray IC emission should not exceed a few tens of MHz \citep[see also][]{sarazin1986}. Given that the X-ray spectral and photon indexes are related as $\Gamma = \alpha_\text{X} + 1$, we fix $\Gamma$ to 1.55 in our \texttt{po} model. The best-fitting normalisation is $(20.7 \pm 2.1) \times 10^{49}$ ph s$^{-1}$ keV$^{-1}$, and the fit quality (reduced C-stat $\simeq 3.8$) is clearly poorer than when using the \texttt{cie} model (Table~\ref{table:fitting_results}, second row).

As a more realistic scenario, one may assume that both thermal and IC processes contribute comparably to the total X-ray emission of this group. Therefore, we refit our spectra with a combination of one \texttt{cie} and one \texttt{po} component (hereafter, the \texttt{cie+po} model). The best-fitting models and their individual components are plotted in Fig.~\ref{fig:spectra}, and the best-fitting parameters are listed in Table~\ref{table:fitting_results} (third row). Compared to the case of a single \texttt{cie} component, the fit does not improve when adding an additional \texttt{po} component and only upper limits on the normalisation of the \texttt{po} component could be obtained ($<1.5 \times 10^{49}$ ph s$^{-1}$ keV$^{-1}$ and $<3.2 \times 10^{49}$ ph s$^{-1}$ keV$^{-1}$ at the 68\% and 90\% confidence levels, respectively accounting for $<11\%$ and $<24\%$ of the total 0.6--5 keV luminosity). For consistency, we also checked the fitting results in MOS (MOS\,1+MOS\,2) and pn separately, as reported in Table~\ref{table:fitting_results} (two last rows). We find an excellent agreement between the MOS and pn best-fitting parameters, always consistent within 1$\sigma$, hence justifying that both instruments can be used and combined to provide more accurate constraints.

\subsection{Possible point-source contamination?}\label{sect:pointsources}

Since the point spread function of the EPIC instruments is not negligible compared to the spatial extent of the source, one may ask whether the observed X-ray emission is contaminated by one or several point sources. To address this question, we inspect the shallow \textit{Chandra}/ACIS observation mentioned in Sect.~\ref{sect:data_xmm} and search for possible point-like sources in the region covered by our EPIC analysis. Among the two source candidates we find, the first one coincides with the centre of the brightest group galaxy while the second has no optical counterpart. Using the CIAO task \texttt{srcflux}, and based on their number of counts, we estimate their 0.5--7 keV fluxes to be $(5\pm 3) \times 10^{-16}$ erg\,s$^{-1}$\,cm$^{-2}$ and $(9\pm 6) \times 10^{-16}$ erg\,s$^{-1}$\,cm$^{-2}$, respectively. If these sources are real, their fluxes do not contribute more than, respectively, $\sim$6\% and $\sim$11\% of the total 0.5--7 keV flux of the diffuse emission measured with EPIC ($\sim 1.4\times 10^{-14}$ erg\,s$^{-1}$\,cm$^{-2}$). Therefore, we conclude that the contamination of the X-ray emitting IGrM by EPIC-unresolved point sources is minimal.

\begin{table*}
\begin{centering}
\caption{Best-fitting parameters and inferred fluxes and luminosities for MRC\,0116+111 (see text). Fixed parameters are indicated with an asterisk (*).}             
\label{table:fitting_results}
\setlength{\tabcolsep}{4pt}
\begin{tabular}{l c c c c c c c c}        
\hline \hline                
Model & $Y_\texttt{cie}$		& $kT$ & $Y_\texttt{po}$	& $\Gamma$ & $f_\text{X,0.5--7}$ & $L_\text{X,0.5--7}$ & $L_\text{X,bol}$ & C-stat/d.o.f.\\    
	   & ($10^{70}$ m$^{-3}$) & (keV) & ($10^{49}$ ph s$^{-1}$ keV$^{-1}$) & & (10$^{-14}$ erg s$^{-1}$ cm$^{-2}$) & (10$^{41}$ erg s$^{-1}$) & (10$^{41}$ erg s$^{-1}$) &\\    
\hline                        
\texttt{cie}	&	$7.6 \pm 0.7$	&	$0.76 \pm 0.07$	&	$-$			&	$-$		&	$1.18 \pm 0.11$	&	$6.4 \pm 0.6$		&	$11.4 \pm 1.0$	&	47.3/36\\
\texttt{po}	&	$-$			&	$-$				&	$20.7 \pm 2.1$	&	$1.55$*	&	$2.7 \pm 0.3$	&	$12.3 \pm 1.2$	&	$165 \pm 17$	&	140.8/37\\
\texttt{cie+po}	&	$7.6 \pm 0.7$	&	$0.76 \pm 0.07$	&	$<1.5$	&	$1.55$*	&	$1.18 \pm 0.11$	&	$6.4 \pm 0.6$	&	$11.4 \pm 1.0$	&	47.3/35\\
\texttt{cie+po} (MOS only)	&  $8.8 \pm 1.2$	&	$0.69 \pm 0.09$	&	$<2.4$	&	$1.55$*	&	$1.33 \pm 0.19$	&	$7.3 \pm 1.0$	&	$13.2 \pm 1.8$	&	27.4/19\\
\texttt{cie+po} (pn only)	&  	$6.8 \pm 1.1$	&	$0.87 \pm 0.12$	&	$<2.4$	&	$1.55$*	&	$1.08 \pm 0.18$	&	$5.8 \pm 0.9$	&	$10.3 \pm 1.7$	&	16.1/13\\
\hline                                   
\end{tabular}
\par\end{centering}
\end{table*}

\subsection{Serendipitous discovery of a rich, distant cluster}\label{sect:newcluster}

\begin{figure}
        \centering
                \includegraphics[width=0.5\textwidth,trim={12 34 12 0},clip]{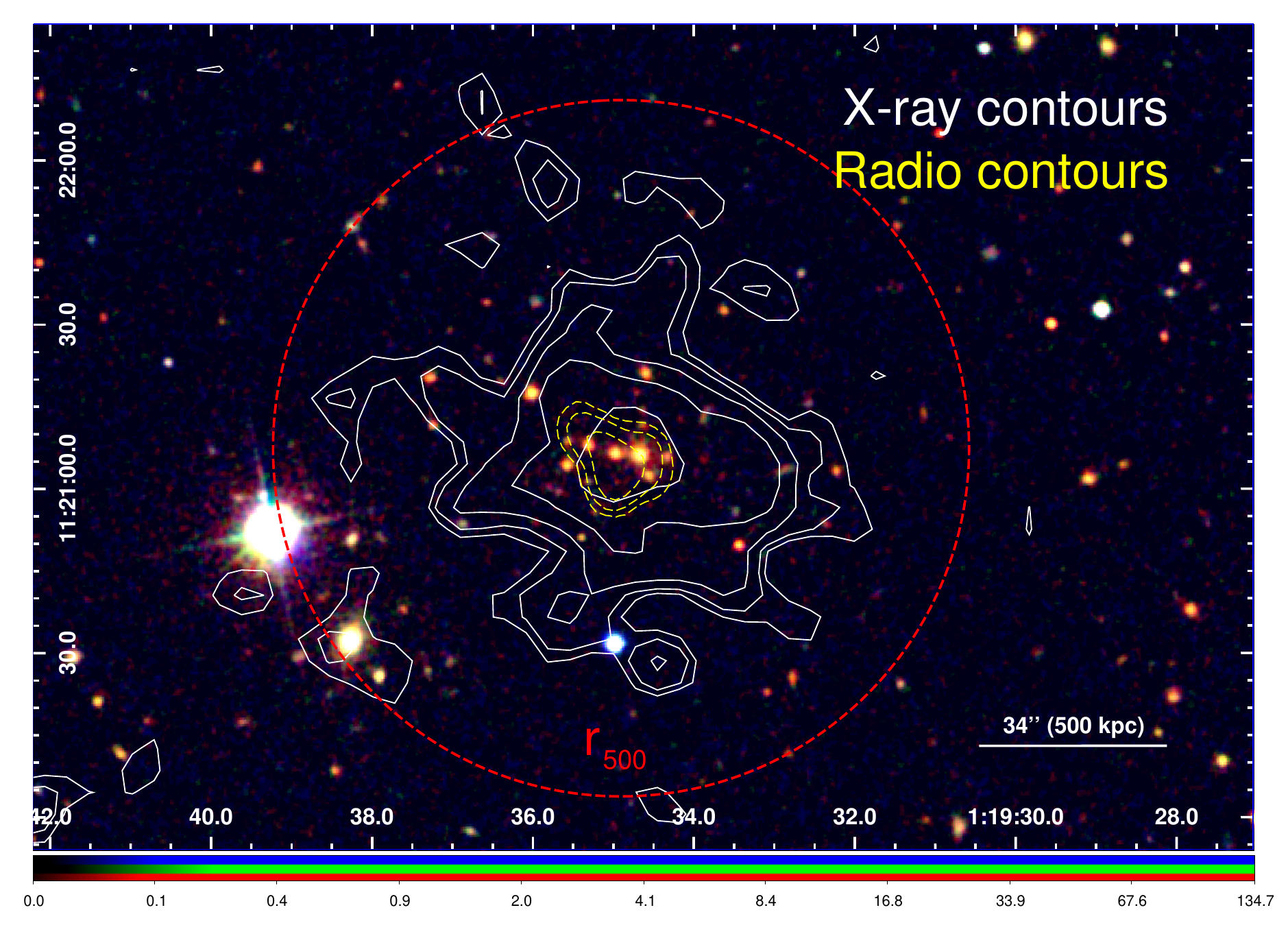}

        \caption{Composite optical image of the serendipitously detected cluster 3XMM\,J011934 from the SDSS-III data ($g$, $r$, and $i$ bands). The X-ray contours (0.3--2 keV, \textit{XMM-Newton} EPIC) are overplotted in white. The radio contours (1.4 GHz, \textit{VLA}) are overplotted in dashed yellow. We also show our best estimate of $r_{500}$ (0.93 Mpc;  dashed red circle).}
\label{fig:3XMM}
\end{figure}

In the field of view of our EPIC pointing, we find an apparent extended emission, at $\sim$1h19$'$34.7$''$ RA, +11$\degr$21$'$06.5$''$ DEC ($\sim$8 arcmin off-axis). 
The source was previously known as a radio emitter, as it was detected in the NVSS \citep[\textit{VLA}, 1.4 GHz;][]{condon1998}, TXS \citep[\textit{Texas Interferometer}, 365 MHz;][]{douglas1996}, TGSS \citep[\textit{GMRT}, 150 MHz;][]{intema2017}, and VLSS \citep[\textit{VLA}, 74 MHz;][]{cohen2007} catalogues. While not detected by \textit{ROSAT} \citep{boller2016}, the X-ray source was formally detected by the 3XMM catalogue \citep[][named as 3XMM\,J011934.7+112106 -- hereafter 3XMM\,J011934]{rosen2016}, although identified as a point-like source. In optical light, an overdensity of galaxies was previously detected in the redMaPPer 6.3.1 catalogue \citep{rykoff2016}, as ID 5105, and with a spectroscopic redshift of $z\simeq 0.525$ for the central galaxy (E. Rykoff, private communication). The source, however, was only labelled as a galaxy cluster candidate because it is situated at the upper redshift limit of the SDSS catalogue. The diffuse X-ray morphology of the source (spanning over at least 1.4 arcmin of diameter and reported for the first time), associated with the optical data described above, provide the decisive evidence for that object being a distant, massive galaxy cluster. An optical RGB mosaic of the source (from the SDSS-III catalogue), overplotted with our EPIC (X-ray) contours and radio contours (1.4 GHz, \textit{VLA}), is shown in Fig.~\ref{fig:3XMM}.

The $r_{500}$ limit of 3XMM\,J011934 can be calculated iteratively by determining $T_X$, i.e. the ICM temperature between $0.1 r_{500} < r < 0.4 r_{500}$, from spectral fits. This radial range is often considered as well representing the "average" cluster temperature \citep[i.e. excluding the possible cool-core and the temperature decrease towards the outskirts; e.g.][]{reiprich2013}. After fixing $r_{500}$ arbitrarily and estimating its corresponding $T_X$, the $r_{500}$ limit can be re-estimated as \citep[][see also Liu et al. \citeyear{liu2018}]{vikhlinin2006}
\begin{equation}
r_{500} = \frac{0.792}{h E(z)} \left( \frac{T_X}{\text{5 keV}} \right)^{0.53} \text{ Mpc,}
\end{equation}
where $h = H_0/100 \equiv 0.7$ and $E(z) = \sqrt{\Omega_m (1+z)^3 + \Omega_\Lambda} \equiv 1.328$. We stop the iteration when $r_{500}$ and $T_X$ provide stable values (i.e. with fluctuations of less than 10\%). We find $r_{500} = 0.93 \pm 0.09$ Mpc, corresponding to $1.06 \pm 0.10$ arcmin\footnote{At $z = 0.525$, 1 arcmin corresponds to $\sim$874 kpc.}.

To determine the dynamical state of this cluster, we estimate the morphological parameter $w$ following the method described in Sect.~\ref{sect:image}. Specifically, we calculate the dispersion of the best-fitting centroids of the source within six concentric apertures spanning from 12 arcsec to 1.1 arcmin ($\sim r_{500}$) and we find $w \simeq 0.019$. When limiting to 500 kpc \citep[i.e. following the prescription of][]{cassano2010}, we find $w \simeq 0.022$, thus not differing much from our first estimate. Similarly, the concentration parameter $c$, defined as the ratio of the measured fluxes within the central 100 and 500 kpc \citep[e.g.][]{cassano2010}, is found to be $c \simeq 0.09$. At first glance, these measurements suggest that 3XMM\,J011934 is a slightly dynamically disturbed, possibly non-cool-core cluster. However, the limited total number of counts ($\sim$2000) might result in spurious structures, which in turn may bias our interpretation. To estimate the typical uncertainties of the $w$ parameter, we add Poisson noise on the best-fitting surface brightness images obtained in each annulus, following the method proposed by \citet{chon2012}. We repeat the exercise 100 times, and we estimate the uncertainty $\Delta w$ as the standard deviation of the $w$ parameters estimated from these mock images. We find $\Delta w = 0.009$ and conclude that the possibility of a relaxed, cool-core cluster cannot be excluded. The effects of the larger point spread function at the off-axis position of the source may also affect the estimate of its morphology \citep{bohringer2010}.

The spatial extent and the number of counts of 3XMM\,J011934 allow to perform basic spatial spectroscopy. Working successively within the circular ranges 0--1 $r_{500}$ (entire cluster) and 0.15--1 $r_{500}$ (core-excised region), we extract and fit the EPIC spectra with a redshifted and absorbed \texttt{cie} model, where the free parameters are the normalisation ($Y_\texttt{cie}$), the temperature ($kT$), and the Fe abundance. The best-fitting parameters are listed in Table~\ref{table:3XMM} and their interpretation is discussed in Sect.~\ref{sect:newcluster_discussion}.

\begin{table}
\begin{centering}
\caption{Best-fitting parameters for 3XMM\,J011934 within $r_{500}$(see text). The central <$0.15 r_{500}$ is successively kept (second column) then excised (last column).}             
\label{table:3XMM}
\setlength{\tabcolsep}{10pt}
\begin{tabular}{l c c}        
\hline \hline                
Parameter & 0--1 $r_{500}$		& 0.15--1 $r_{500}$ \\    
\hline                        
$Y_\texttt{cie}$	($10^{72}$ m$^{-3}$)	 &	$9.3 \pm 1.4$	&	$7.8 \pm 1.4$\\
$f_\text{X,bol}$	($10^{-13}$ erg\,s$^{-1}$\,cm$^{-2}$)	 &	$1.48 \pm 0.22$	&	$1.32 \pm 0.23$\\
$L_\text{X,bol}$	($10^{44}$ erg\,s$^{-1}$) 		 &	$2.1 \pm 0.3$	&	$1.8 \pm 0.3$\\
$kT$	 (keV)					 	 &	$4.4_{-0.5}^{+0.7}$	&	$4.5 \pm 0.7$\\
Fe	 							 &	$1.1_{-0.6}^{+0.8}$	&	$1.3_{-0.7}^{+1.1}$\\
\hline
C-stat/d.o.f. 						 &	$98.3/58$	&	$104.0/58$\\

\hline                                   
\end{tabular}
\par\end{centering}
\end{table}


\section{Discussion}\label{sect:discussion}


\subsection{Inverse-Compton emission and volume-averaged magnetic field}\label{sect:IC_B}

Presumably, both the synchrotron radiation observed in diffuse radio sources and the CMB photons that are boosted to X-ray energies via IC scattering originate from the same population of relativistic electrons. Because the radio synchrotron emission depends on both the relativistic electron distribution and the volume-averaged magnetic field while the IC emission depends on the electron distribution only, constraining radio and X-ray IC fluxes allows us to put further constraints on the average magnetic field of a galaxy cluster/group. Quantitatively, and assuming the relativistic electron population to be distributed as a function of their Lorentz factor $\gamma$ as $N(\gamma) = N_0 \gamma^{-p}$, the radio and X-ray fluxes $f_r(\nu_\text{syn})$ and $f_x(\nu_\text{IC})$, emitting, respectively, at the frequencies $\nu_\text{syn}$ and $\nu_\text{IC}$, can be written \citep[eqs 4.59 and 2.65 of][see also Ota et al. \citeyear{2014A&A...562A..60O}]{1970RvMP...42..237B}:
\begin{equation}
f_r(\nu_\text{syn}) \equiv \frac{dW_\text{syn}}{d\nu_\text{syn} dt} = \frac{4\pi N_0 e^3 B^\frac{p+1}{2}}{m_e c^2} \left(\frac{3e}{4\pi m_e c}\right)^\frac{p-1}{2} a(p) \, \nu_\text{syn}^{-\frac{p-1}{2}}
\end{equation}
\begin{equation}
f_x(\nu_\text{IC}) \equiv \frac{dW_\text{IC}}{d\nu_\text{IC} dt} = \frac{8\pi^2 N_0 r_0^2}{c^2} h^{-\frac{p+3}{2}} (kT_\text{CMB})^\frac{p+5}{2} F(p) \, \nu_\text{IC}^{-\frac{p-1}{2}}
\end{equation}
where $B$ is the volume-averaged magnetic field, $e$ and $m_e$ are, respectively, the electron charge and mass, $r_0 = \frac{e^4}{m_e^2 c^4}$ is the classical electron radius, $h$ is the Planck constant, and $T_\text{CMB} = 2.73 (1+z)$ is the average temperature of the CMB at the considered redshift. The functions $a(p)$ and $F(p)$ depend on the gamma function $\Gamma(x)$ and the Riemann zeta function $\zeta(x)$ as
\begin{equation}
a(p) = 2^\frac{p-7}{2} \sqrt{\frac{3}{\pi}}\, \frac{\Gamma\left(\frac{3p-1}{12}\right) \Gamma\left(\frac{3p+19}{12}\right) \Gamma\left(\frac{p+5}{4}\right)}{(p+1) \, \Gamma\left(\frac{p+7}{4}\right)}
\end{equation}
\begin{equation}
F(p) = 2^{p+3}\frac{p^2 + 4p + 11}{(p+3)^2 (p+1) (p+5)} \Gamma\left(\frac{p+5}{2}\right) \zeta\left(\frac{p+5}{2}\right).
\end{equation}
The ratio between radio and X-ray fluxes at the two given frequencies provides thus \citep[eq. 5.10 of][]{sarazin1986}:
\begin{equation}
B^\frac{p+1}{2} = \frac{2\pi e h^{-\frac{p+3}{2}}}{m_e c^4}  \left(\frac{4\pi m_e c}{3e}\right)^\frac{p-1}{2} \frac{f_r(\nu_\text{syn})}{f_x(\nu_\text{x})} \left(\frac{\nu_\text{syn}}{\nu_\text{IC}}\right)^\frac{p-1}{2} \frac{F(p)}{a(p)} (kT_\text{CMB})^\frac{p+5}{2}.
\end{equation}

Since the synchrotron spectral index and the slope of the power-law electron distribution are related as $\alpha_\text{syn} = \frac{p-1}{2}$, the radio spectral index of $\alpha_\text{syn} = 0.55$ \citep{bagchi2009} translates into $p = 2.1$. At 2 keV, the combined EPIC instruments provide a flux upper limit of <1.53 $\times 10^{-10}$ Jy (<3.34 $\times 10^{-10}$ Jy at 90\% confidence). Adopting the radio flux estimate of 1.15 Jy at 240 MHz \citep{bagchi2009}, we find that, at 68\% confidence, the average magnetic field is higher than $\sim$4.3 $\mu$G (higher than $\sim$2.6 $\mu$G at 90\% confidence). We note that, in the unlikely case of the observed X-ray emission being significantly contaminated by point sources (Sect.~\ref{sect:pointsources}), the upper limit on the IC flux would be lower, and thus the lower limits reported above would be even higher. Repeating the same exercise within the north-western lobe only (assuming that the X-ray emission -- detected as an upper limit only -- originates entirely from IC scattering), we find a lower limit for the average magnetic field of $\gtrsim$2.0 $\mu$G.

Volume-averaged magnetic fields of the same order have been previously reported by \citet{croston2005} by comparing the radio and X-ray emission in the radio lobes of isolated radio-galaxies and quasars. The present case is different, as the radio (and X-ray) emission is not clearly concentrated in lobes or jets around one galaxy, but rather widespread over an entire galaxy group. In fact, this is the first time to our knowledge that such a high lower limit of the volume-averaged magnetic field is reported in the case of galaxy groups and/or clusters with this method. 
Instead, previous measurements obtained in radio haloes of clusters using the same method estimated lower limits of typically 0.1--1 $\mu$G \citep[][and references therein]{bartels2015}, whereas, interestingly, \citet{bagchi1998} obtained a volume-averaged magnetic field of $\sim$1 $\mu$G in the central steep-spectrum radio relic region of the rich galaxy cluster Abell\,85, i.e. at least a factor of 2 lower than our estimate. Other methods -- either Faraday rotation measurements \citep[e.g.][]{bohringer2016} or the assumption of equipartition \citep[e.g.][]{feretti2012} -- provide higher estimates of cluster/group magnetic fields, at first glance consistent with our measurements. However, such comparisons are very limited because (i) Faraday rotation measurements depend on the magnetic fields along the line-of-sight and are thus very sensitive to the magnetic field topology in clusters/groups, and (ii) the assumption of equipartition remains questionable given that relativistic electrons, exhibiting a relatively short lifetime, and magnetic fields, associated with the much longer lifetimes of clusters, might have different origins \citep[e.g.][]{carilli2002,petrosian2008}.

\subsection{An extremely powerful past AGN activity?}\label{sect:AGN_activity}

\begin{figure}
        \centering
                \includegraphics[width=0.5\textwidth]{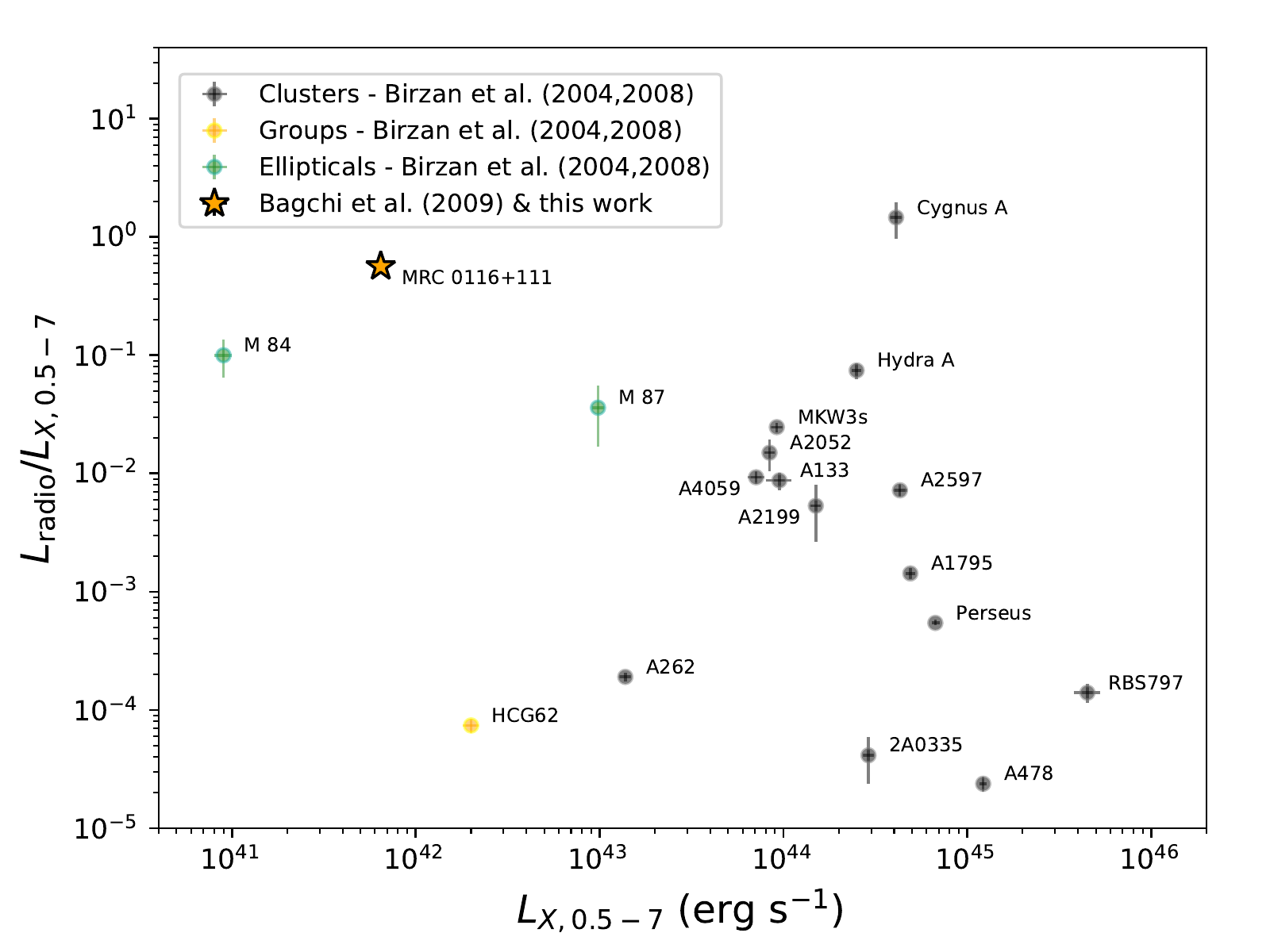}
        \caption{Ratio between the radio luminosity over the range 10 MHz--10 GHz ($L_\text{radio}$) and the X-ray luminosity in the 0.5--7 keV band ($L_\text{X,0.5--7}$), plotted for MRC\,0116+111 and other clusters, groups, and ellipticals \citep{birzan2004,birzan2008}.}
\label{fig:L_radio_xray}
\end{figure}

Besides the surprisingly high lower limit of its volume-averaged magnetic field, MRC\,0116 is also a unique system in terms of radio vs. X-ray emission. In Fig.~\ref{fig:L_radio_xray}, we compare the ratio $L_\text{radio}/L_\text{X,0.5--7}$ (namely the radio luminosity integrated between 10 MHz and 10 GHz over the X-ray luminosity in the 0.5--7 keV band) of our source with that of several other systems \citep{birzan2004,birzan2008}. Quite remarkably, such a radio-to-X-ray luminosity ratio for MRC\,0116 is almost 70 times higher than the median ratio of all the other systems. In particular, it is $\sim$13 times and almost $\sim$5 times stronger than respectively M\,87 and M\,84. These two elliptical galaxies, however, are both known for hosting powerful, still active synchrotron jets launched from a bright point-like radio source coinciding with the position of the AGN \citep[e.g.][]{owen2000,laing2011}. These bright radio features are not found in MRC\,0116 that, instead, shows a much more diffuse and extended emission \citep[see the higher resolution -- beam size 5 arcsec --\textit{GMRT} 1.28 GHz map in figure 1 of][]{bagchi2009}. From the sample of \citet{birzan2004}, Cygnus A is the only system exhibiting a higher $L_\text{radio}/L_{X}$ ratio than MRC\,0116. However, the bulk of the radio emission in Cygnus A originates from its nucleus and hotspots, rather than from the surrounding diffuse lobes \citep[e.g.][]{carilli1991,mckean2016}. This makes MRC\,0116, to our knowledge, the brightest radio vs. X-ray extragalactic \emph{diffuse} source reported so far.

\begin{figure}
        \centering
                \includegraphics[width=0.5\textwidth]{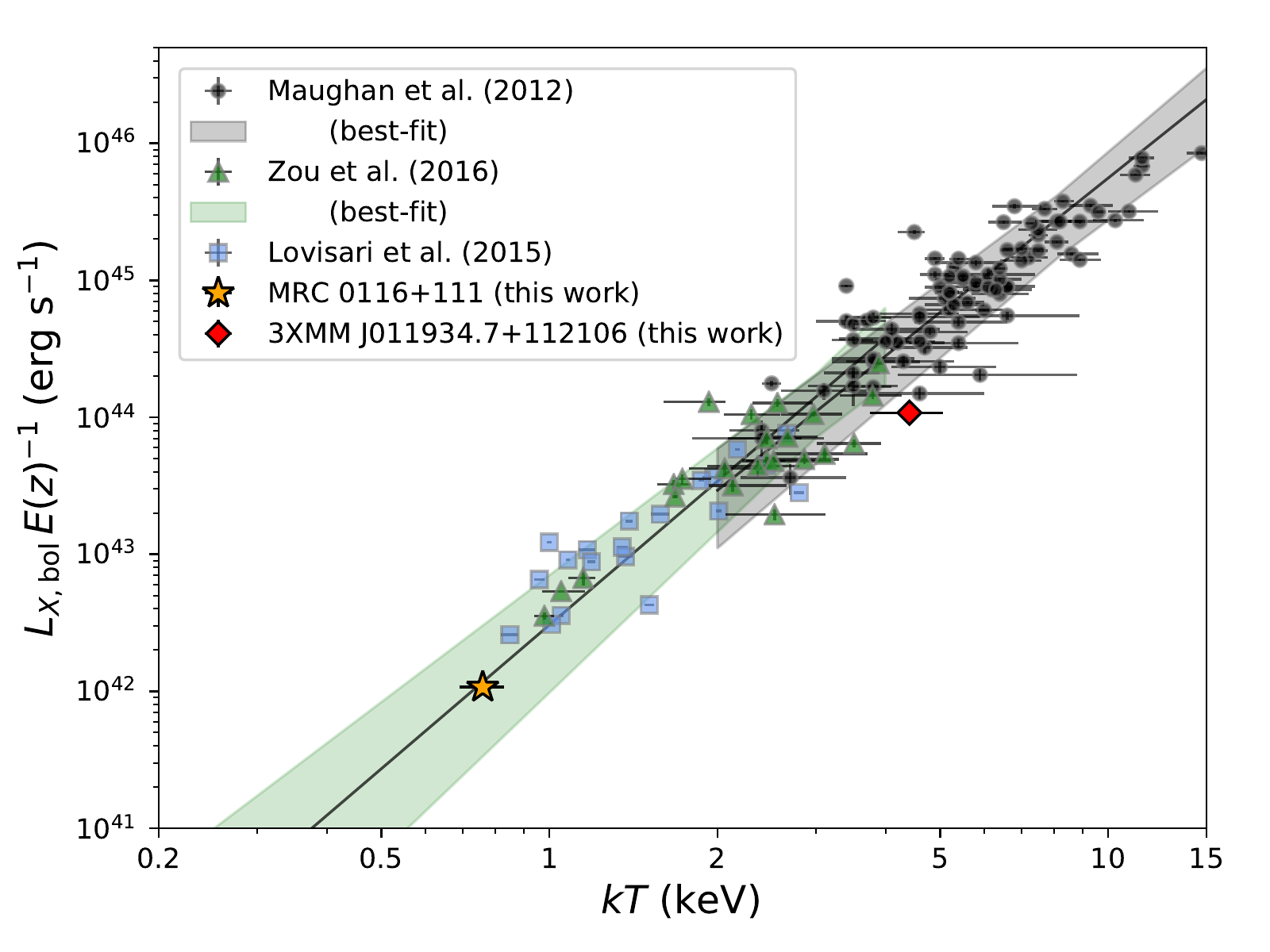}

        \caption{Temperature--bolometric luminosity relation comparing several groups and clusters previously measured \citep{maughan2012,zou2016} with our estimates for MRC\,0116+111 (the yellow star; Sect.~\ref{sect:AGN_activity}) and 3XMM\,J011934 (the red diamond; Sect.~\ref{sect:newcluster_discussion}). Measurements and the best-fitting relation from \citet{maughan2012} and \citet{zou2016} are adopted for unrelaxed systems (respectively, classified as "$\langle w \rangle \ge 0.006$" and "non-relaxed cool-core") without core excision. For comparison, galaxy groups measurements from the sample of \citet{lovisari2015} are also plotted (where the bolometric luminosity is extrapolated from the luminosity in the 0.1--2.4 keV band, see text).}
\label{fig:L_T}
\end{figure}

A question of interest is whether such an unusually high ratio is due to one (or several) intense past AGN outburst(s) that had severely disturbed its surrounding IGrM and expelled a significant fraction of the thermal gas outside its gravitational well. Presumably, such a hypothetical outburst would have reduced the overall X-ray luminosity of the system, while it would have provided a substantial source of (re-)heating to the remaining fraction of the radiatively cooling gas, thereby shifting the source away from the well-established $L_X$--$T$ scaling relation (tracing the self-similar evolution of groups and clusters). Such a scaling relation, taken from recent observations of (mostly unrelaxed) clusters \citep{maughan2012} and groups \citep{lovisari2015,zou2016}, and its comparison with the corresponding measurements obtained for MRC\,0116, is plotted in Fig.~\ref{fig:L_T}. Since we adopt $L_X$ as the X-ray bolometric luminosity \citep[$L_{X\text{, bol}}$;][and their corresponding best-fitting relations]{maughan2012,zou2016}, the luminosities from \citet{lovisari2015} given in the 0.1--2.4 keV band are extrapolated to their bolometric counterpart by calculating appropriate redshifted, absorbed \texttt{cie} models in \textsc{SPEX}. As shown in Fig.~\ref{fig:L_T}, the comparison results in an excellent agreement between our X-ray luminosity and temperature measurements of MRC\,0116 and the best-fitting relation from recent estimates for other systems, with no excess of $kT$ nor deficiency of $L_{X\text{, bol}}$. This suggests that the very high $L_\text{radio}/L_{X}$ ratio is explained rather by unusually bright radio emission than by unusual X-ray gas properties. In fact, although MRC\,0116 is a poor group, its radio luminosity $L_\text{radio}$ \citep[$3.64 \times 10^{41}$ erg s$^{-1}$;][]{bagchi2009} is comparable to that of M\,87 \citep[with a rather different emission distribution, as almost 50\% of the flux of the latter originates from its nucleus;][]{owen2000} and is several factors more than those found in mini-haloes of much richer clusters \citep[e.g. MKW\,3s, A\,262, 2A\,0335+096, A\,478, Centaurus;][]{birzan2008}.

More generally, one may ask about the true nature of such a radio synchrotron emission exceptionally bright for a group like MRC\,0116. Based on the different types of diffuse radio sources typically found at the centre of large-scale, non-merging systems, potential candidates would be (i) a "mini-halo"-like source or (ii) a series of radio bubbles related to past and recent AGN activity. The radio morphology of the source is rather similar to mini-haloes found at the centre of, e.g. Perseus \citep{gendron2017}; however the extent of the radio source -- usually limited to cluster cores in the case of mini-haloes -- is about three times larger than the extent of the observed X-ray emission. Moreover, mini-haloes are typically found in and associated with relaxed systems \citep[e.g.][]{gitti2004}, which might not be necessarily the case for MRC\,0116 (Sect.~\ref{sect:image}; see also Figs. \ref{fig:xray_radio_image} and \ref{fig:xray_radio_contours} left). On the other hand, in addition to the absence of a point-like radio source at the position of the central dominant galaxy \citep{bagchi2009}, the bulk of the radio source appears too diffuse and too extended to be entirely attributed to a series of well-separated bubbles witnessing of past episodic AGN outbursts. While the two inner lobe-like structures may definitely be related to jets of a relativistic plasma from a recent outburst of the central SMBH (as traced by the possible eastern X-ray cavity reported in Sect.~\ref{sect:image}), the surface brightness of the outer parts of the radio source remains rather uniform \citep[Figs. \ref{fig:xray_radio_image} and \ref{fig:xray_radio_contours} right; see also the higher resolution \textit{GMRT} 1.28 GHz map in][]{bagchi2009}.

It is possible, however, that the large-scale radio emission originates from an older population of relativistic electrons, initially seeded by buoyantly rising bubbles of non-thermal plasma (extending towards north-west and south-east), and later re-energized by turbulent motions, likely triggered by a more recent outburst. This hypothesis would also naturally explain the sudden ageing of electrons in the north-western lobe -- i.e. where the radio-emitting plasma extends beyond the relatively dense central IGrM -- as well as the uniformity of the radio spectral index in regions covered by the X-ray emitting gas \citep[see figure 6 right in][]{bagchi2009}. If this is the case, MRC\,0116 would be a unique system, hosting a disturbed hot atmosphere, in which a remarkably powerful past AGN activity would have injected enough turbulence and small-scale motions to re-accelerate relativistic electrons, though without significantly affecting the overall temperature or the gas density of the system. Deeper \textit{XMM-Newton} and/or \textit{Chandra} observations, as well as future X-ray missions with micro-calorimeters onboard (which would allow direct measurements of the widths and resonance scattering of Fe-L lines), will be necessary to further confirm or rule out this hypothesis.

\subsection{Basic properties of 3XMM\,J011934}\label{sect:newcluster_discussion}

When compared with the $L_X$--$T$ relation of nearby systems (Fig.~\ref{fig:L_T}, see also Table~\ref{table:3XMM}), the serendipitously discovered cluster 3XMM\,J011934 lies within the scatter obtained in previous studies. We note, however, that the cluster appears somewhat underluminous (and/or hotter) compared to the self-similar relation fitted by \citet{maughan2012}. Assuming that such a trend is real and entirely due to the higher redshift of the source ($z\simeq 0.525$), it would agree qualitatively with the observational results of \citet{reichert2011}, who found hints for distant clusters to be less luminous than their self-similar prediction. On the other hand, these results are still under debate, as a steepening of the $L_X$--$T$ relation with redshift was recently supported by observations \citep{giles2016} and numerical simulations \citep{truong2018}. Clearly, more high-$z$ systems are needed to confirm and interpret a possible evolution of the $L_X$--$T$ relation with cosmic time. Future missions such as \textit{eROSITA} and \textit{Athena} will be essential for this purpose.

Another interesting feature of 3XMM\,J011934 is its Fe abundance consistent with solar in both our entire and core-excised extracted regions. Whereas a solar value is commonly found in the cool cores of nearby, relaxed systems \citep[for a review, see][]{mernier2018} the Fe abundance that we measure outside 0.15 $r_{500}$ is $>$1$\sigma$ higher than the mean estimate in similar regions of other clusters at $z \simeq 0.5$ \citep{mcdonald2016,mantz2017}. Admittedly, our statistical uncertainties are still large and deeper re-observation of 3XMM\,J011934 would be necessary to get better observational constraints on its chemical enrichment.


\section{Conclusions}\label{sect:conclusion}


Using the EPIC instruments onboard \textit{XMM-Newton}, we have presented for the first time a detailed X-ray observation of the poor galaxy group MRC\,0116+111 ($z=0.132$). This system hosts a bright, diffuse radio emission \citep{bagchi2009}. The radio morphology and the remarkably high radio-to-X-ray luminosity (to our knowledge the highest for a diffuse extragalactic source) strongly suggest that the group has experienced an intense AGN activity from its central cD galaxy over the last $\sim$100 Myr. Although the thermal X-ray emitting IGrM appears morphologically disturbed and is about three times less extended in projection than the non-thermal radio-emitting plasma, this source is not found to deviate from the $L_X$--$T$ scaling relation established for more massive groups and clusters. This suggests that, despite its power, the AGN jets/lobes were not efficient at ejecting baryons nor at heating the thermal IGrM substantially. Instead, a past outburst may have efficiently stirred the hot atmosphere, possibly translating into turbulent re-acceleration of a significant fraction of relativistic electrons traced by the radio emission. This scenario would qualitatively explain the overlap of the region of flat radio spectral index with that of the X-ray emitting gas.

Because of the relatively low temperature of less than 1 keV, only negligible thermal X-ray emission is expected in the 2--10 keV band, this source is an ideal target to search for and constrain the emission originating from IC scattering of the CMB photons by the relativistic electron population. Although the limited cleaned exposure allows only to derive an upper limit for this non-thermal X-ray flux, this also translates into a lower limit for the group's volume-averaged magnetic field of $\ge$4.3 $\mu$G. Although this is not inconsistent with the typical magnetic field estimates derived using the assumption of equipartition and/or Faraday rotation measurements, this is, to our knowledge, the highest lower limit reported to date using constraints from the IC X-ray emission. Deeper \textit{XMM-Newton} observations and future missions will help setting better constraints on its average magnetic field intensity and on the coupling of the AGN feedback with the thermal and non-thermal components of the plasma.

In the field of view of our EPIC observation, we also serendipitously discovered a distant ($z\simeq 0.525$) galaxy cluster. The X-ray emission, which had been previously detected as a point-like source in the 3XMM catalogue (ID: 3XMM\,J011934.7+112106), clearly shows an extended morphology, and its location is coincident with an overdensity of galaxies, reported by the redMaPPer 6.3.1 catalogue but labelled as a cluster candidate only. Our analysis suggests a moderately hot ($\sim 4.5$ keV), possibly unrelaxed cluster, which also exhibits radio emission in its core.

\section*{Acknowledgements}

The authors thank the anonymous referee for insightful comments that helped to improve this paper, as well as Nhut Truong, Eli S. Rykoff, Lorenzo Lovisari, Esra Bulbul, Soumyajit Mandal, Grant Tremblay, Gerrit Schellenberger, and Vittorio Ghirardini for fruitful discussions. This work is supported by the Lend\"ulet LP2016-11 grant awarded by the Hungarian Academy of Sciences. JJ would like to acknowledge the support received from IUCAA, Pune, India as a visiting Associate there. AS is supported by the Women In Science Excel (WISE) programme of the Netherlands Organisation for Scientific Research (NWO), and acknowledges the MEXT World Premier Research Center Initiative (WPI) and the Kavli IPMU for the continued hospitality. Funding for SDSS-III has been provided by the Alfred P. Sloan Foundation, the Participating Institutions, the National Science Foundation, and the U.S. Department of Energy Office of Science. The SDSS-III web site is http://www.sdss3.org/. SDSS-III is managed by the Astrophysical Research Consortium for the Participating Institutions of the SDSS-III Collaboration including the University of Arizona, the Brazilian Participation Group, Brookhaven National Laboratory, Carnegie Mellon University, University of Florida, the French Participation Group, the German Participation Group, Harvard University, the Instituto de Astrofisica de Canarias, the Michigan State/Notre Dame/JINA Participation Group, Johns Hopkins University, Lawrence Berkeley National Laboratory, Max Planck Institute for Astrophysics, Max Planck Institute for Extraterrestrial Physics, New Mexico State University, New York University, Ohio State University, Pennsylvania State University, University of Portsmouth, Princeton University, the Spanish Participation Group, University of Tokyo, University of Utah, Vanderbilt University, University of Virginia, University of Washington, and Yale University. This work is based on observations obtained with \textit{\textit{XMM-Newton}}, an ESA science mission with instruments and contributions directly funded by ESA member states and the USA (NASA). The scientific results reported in this article are based in part on data obtained from the \textit{Chandra} Data Archive. SRON is supported financially by NWO.




\bibliographystyle{mnras}
\bibliography{MRC0116}




 \appendix
 
 \section{Background-related uncertainties}\label{sect:bg_uncertainties}
 
As mentioned in Sect.~\ref{sect:data_xmm}, the relatively weak source-to-background ratio of our observation of MRC\,0116 makes our results sensitive to the background determination. Our approach of subtracting a local background region (located on the same detector chip as the source) seems reliable in our case, mainly because (i) time variation of the non-X-ray (i.e. hard particle and remaining soft proton) background is synchronized between the "source" and the "background" regions and (ii) the X-ray foreground (local hot bubble and Galactic thermal emission) is not expected to change significantly at such small angular scales. For safety, we also select an alternative, less extended background region and we verify that the results presented throughout this paper are not significantly affected by this change. Nevertheless, it is important to verify that our results remain consistent also when adopting another background treatment.

An alternative reliable method commonly adopted in the literature is to model all the background components directly in the spectra of the source. We choose to do so following the general method described extensively in \citet[][and references therein]{mernier2015}. In summary, we extract EPIC spectra covering the entire field of view of our observation (after discarding all the detected point-like sources and the regions covering MRC\,0116 and 3XMM\,J011934). Although, admittedly, the spectral shape of some components may somewhat vary with their position on the sky/detector, such a large extraction region is necessary to provide accurate constraints on the average behaviour of the remaining soft proton background (expected to be non-negligible in our case, as the observation was highly flared). The local hot bubble and Galactic thermal emission are modelled respectively by an unabsorbed and absorbed \texttt{cie} model, assuming proto-solar abundances, $kT_\text{LHB} = 0.08$ keV, and $kT_\text{GTE}$ left free. The cosmic X-ray background is modelled by an absorbed power-law ($\Gamma_\text{CXB} = 1.41$) and its normalisation is fixed to match the integrated flux expected below the faintest detected point-like source \citep[estimated using the tool \texttt{cxbups};][]{mernier2015,deplaa2017}. The remaining soft proton background and the hard particle background, both unfolded by the effective area, are modelled respectively by a power-law with free photon index $0.1 < \Gamma_\text{SP} < 1.4$ and by a broken power-law fitted from EPIC filter wheel closed data -- using Gaussian functions with free normalisation to reproduce the instrumental lines. We then adopt and re-scale these background components to the source spectra, also taking vignetting effects into account in the case of the remaining soft proton background.

Modelling the source spectra with a \texttt{cie+po} combination in addition to the above background components, we find $kT_\texttt{cie} = (0.63 \pm 0.11)$ keV and $Y_\texttt{po} < 2.7 \times 10^{49}$ ph s$^{-1}$ keV$^{-1}$, thus formally consistent with our previous estimates. This translates into a 1$\sigma$ volume-averaged magnetic field lower limit of $\gtrsim$2.9 $\mu$G. This limit is somewhat lower than the one reported in Sect.~\ref{sect:IC_B} (4.3 $\mu$G), probably because modelling the background naturally provides higher uncertainties, however it remains higher than the 90\% confidence value of the latter (2.6 $\mu$G). We conclude that background-related uncertainties are under control, as they can be fairly covered by adopting a 2$\sigma$ confidence interval for the volume-averaged magnetic field lower limit. Even though subtracting the background (Sect. \ref{sect:data_xmm}) is certainly the safest approach in this case, the fact that the $\gtrsim$2.9 $\mu$G lower limit inferred in this section remains remarkably high compared to previous values from the literature confirms the surprising results on MRC\,0116 reported and discussed above.


\bsp	
\label{lastpage}
\end{document}